\documentclass[sigconf]{acmart}
\usepackage{multirow}
\usepackage{graphicx}
\usepackage{subcaption}
\usepackage{hyperref}
\usepackage{xcolor}
\definecolor{participantblue}{RGB}{80, 130, 200}   % light-medium blue for text & border
\definecolor{lightbluebg}{RGB}{235, 245, 255}     % soft blue background

\newenvironment{participantquote}
  {\begin{quote}\itshape}
  {\end{quote}}

\setcopyright{none}
\renewcommand\footnotetextcopyrightpermission[1]{} % removes footnote with conference information in first column
\AtBeginDocument{%
  }

\copyrightyear{2026}
\acmYear{2026}
\setcopyright{cc}
\setcctype{by}
\acmConference[CAIN '26]{2026 IEEE/ACM 5th International Conference on AI Engineering - Software Engineering for AI}{April 12--13, 2026}{Rio de Janeiro, Brazil}
\acmBooktitle{2026 IEEE/ACM 5th International Conference on AI Engineering - Software Engineering for AI (CAIN '26), April 12--13, 2026, Rio de Janeiro, Brazil}
\acmDOI{10.1145/3793653.3793775}
\acmISBN{979-8-4007-2475-6/2026/04}

\begin{document}

\title{%Dealing with Data Leakage in Automotive Perception Functions Development: Practitioners' Perspectives
%Understanding Data Leakage in Automotive Perception: Insights from Practitioners
Data Leakage in Automotive Perception: Practitioners' Insights}


\author{Md Abu Ahammed Babu}
\affiliation{%
  \institution{Volvo Cars | University of Gothenburg and Chalmers University of Technology}
  \city{Gothenburg}
  \country{Sweden}
}\email{md.abu.ahammed.babu@volvocars.com}

\author{Sushant Kumar Pandey}
\affiliation{%
\institution{University of Groningen}
  \city{Groningen}
  \country{Netherlands}} \email{s.k.pandey@rug.nl}

\author{Darko Durisic}
\affiliation{%
  \institution{Volvo Cars}
  \city{Gothenburg}
  \country{Sweden}
}\email{darko.durisic@volvocars.com}

\author{András Bálint}
\affiliation{%
  \institution{Volvo Cars}
  \city{Gothenburg}
  \country{Sweden}
}\email{andras.balint@volvocars.com}

\author{Miroslaw Staron}
\affiliation{%
  \institution{University of Gothenburg and Chalmers University of Technology}
  \city{Gothenburg}
  \country{Sweden}}
\email{miroslaw.staron@gu.se}

\renewcommand{\shortauthors}{Babu et al.}
%%
%% Article type: Research, Review, Discussion, Invited or position
\acmArticleType{Review}
%%
%% Links to code and data
%\acmCodeLink{https://github.com/borisveytsman/acmart}
%\acmDataLink{htps://zenodo.org/link}
%%
%% Authors' contribution
% \acmContributions{BT and GKMT designed the study; LT, VB, and AP
%   conducted the experiments, BR, HC, CP and JS analyzed the results,
%   JPK developed analytical predictions, all authors participated in
%   writing the manuscript.}
%%
%% Sometimes the addresses are too long to fit on the page.  In this
%% case uncomment the lines below and fill them accodingly.
%%
% \authorsaddresses{Corresponding author: Md Abu Ahammed Babu,
% \href{mailto:md.abu.ahammed.babu@volvocars.com}{md.abu.ahammed.babu@volvocars.com};
% Volvo Cars, Gothenburg, Sweden}
%%
%%
\begin{abstract}
    Data leakage is the inadvertent transfer of information between training and evaluation datasets that poses a subtle, yet critical, risk to the reliability of machine learning (ML) models in safety-critical systems such as automotive perception. While leakage is widely recognized in research, little is known about how industrial practitioners actually perceive and manage it in practice. This study investigates practitioners' knowledge, experiences, and mitigation strategies around data leakage through ten semi-structured interviews with system design, development, and verification engineers working on automotive perception functions development. Using reflexive thematic analysis, we identify that knowledge of data leakage is widespread and fragmented along role boundaries: ML engineers conceptualize it as a data-splitting or validation issue, whereas design and verification roles interpret it in terms of representativeness and scenario coverage. Detection commonly arises through generic considerations and observed performance anomalies rather than implying specific tools. However, data leakage prevention is more commonly practiced, which depends mostly on experience and knowledge sharing. These findings suggest that leakage control is a socio-technical coordination problem distributed across roles and workflows. We discuss implications for ML reliability engineering, highlighting the need for shared definitions, traceable data practices, and continuous cross-role communication to institutionalize data leakage awareness within automotive ML development.
\end{abstract}


\maketitle

\section{Introduction}
\label{sec: introduction}
Artificial intelligence (AI) and machine learning (ML) have become integral to modern automotive software systems, underpinning critical functionalities such as object detection, driver monitoring, and autonomous navigation \cite{kandregula2020exploring}. As vehicles increasingly rely on perception models to make decisions in dynamic environments, the reliability of these models becomes not merely a matter of accuracy but also of consistent and dependable operation \cite{rosique2019systematic}. The performance of ML components, however, depends heavily on the integrity of the data they are trained and evaluated on. Small and often overlooked errors in data handling, such as improper dataset splitting (i.e., separating the training and evaluation data), inadvertent duplication, or inclusion of future information during training, can lead to \emph{data leakage}, a phenomenon where the model “learns” information that should not be available during the training process. This relatively simple issue might produce misleadingly high validation results while concealing poor generalization leading to suboptimal real-world performance.

Data leakage has long been recognized as one of the most insidious threats to trustworthy ML systems. Studies have documented its occurrence across domains—from healthcare diagnostics to autonomous driving—often leading to inflated performance scores and compromised decision-making \cite{kapoor2023leakage}. In computer vision, leakage may occur through overlapping frames, correlated scenes, or metadata patterns that persist across training and test sets \cite{heyn2023automotive}. Within automotive perception specifically, even publicly released benchmark datasets have shown subtle forms of data leakage due to temporal or spatial dependencies that violate the independence of evaluation data \cite{babu2025d}. Such issues, though well-known in theory, remain difficult to detect in complex industrial pipelines where multiple actors contribute to data creation, annotation, development, and testing \cite{sasse2025overview}.

Despite the rising attention to responsible AI engineering, empirical studies examining how data leakage is understood, detected, and mitigated in practice remain scarce. Prior work in software engineering for ML has focused largely on technical challenges such as data and model versioning \cite{amershi2019software}, reproducibility and documentation \cite{serban2020adoption}, and pipeline automation \cite{serban2024software}, while organizational and human factors have received less attention. Even in safety-conscious sectors such as automotive, existing frameworks like ISO~26262 \cite{ISO26262} and SOTIF (ISO/PAS~21448) emphasize functional safety and system verification, yet offer limited guidance on ensuring data integrity within ML workflows \cite{borg2020safely}. Consequently, awareness of data leakage often depends on individual experience rather than formal process standards, and practices for detection or prevention are inconsistently applied across teams.

Moreover, industrial ML development rarely involves a single homogeneous group. Functions related to internal or external perception typically span several roles, including system design engineers, ML developers, data scientists, and verification engineers — each contributing to different stages of the software development life cycle (SDLC). While ML specialists may recognize data leakage as a technical pitfall, upstream design engineers or downstream verification engineers may perceive it only indirectly through system behavior. Understanding these diverse perspectives is essential for developing cross-role safeguards that align with the realities of industrial ML engineering.

In this paper, we present an exploratory case study conducted at an automotive OEM (Original Equipment Manufacturer) through interviewing ten practitioners from two automotive software development teams responsible for developing ML-enabled functionalities: one working with internal perception functions development and the other developing external perception-related functions. Participants represented a range of roles within the SDLC, from system design and architecture to ML model development and verification. The study investigates how practitioners define and perceive data leakage across roles, what experiences they have had with it, and what measures they adopt to detect and prevent it in practice. Our goal is to understand the state of knowledge, existing practices, and organizational conditions that influence the management of data leakage risks in automotive software development.

To guide our research, we formulated the following research questions:

\begin{enumerate}
  \item[] \textbf{RQ1:} What knowledge do practitioners have across different roles about data leakage and related risks in ML-based automotive software development?  
  \item[] \textbf{RQ2:} How have practitioners experienced data leakage in their work, and what were the perceived impacts and responses to such occurrences?  
  \item[] \textbf{RQ3:} What practices are used by teams to detect and prevent data leakage in their workflows?  
  \item[] \textbf{RQ4:} What are the recommended guidelines suggested by practitioners to mitigate risks of data leakage in current development practices?  
\end{enumerate}

By analyzing insights from these interviews, we aim to contribute a grounded understanding of data leakage knowledge and management across diverse engineering roles. The study highlights how data-centric reliability concerns intersect with established safety and software development processes. Ultimately, our findings aim to inform guidelines for data management and ML assurance practices in the automotive domain.

The remainder of this paper is structured as follows. Section~\ref{sec: background} reviews related work on data leakage and ML engineering. Section~\ref{sec: methodology} details our study design and analysis approach. Section~\ref{sec: findings} presents the key findings organized around the research questions. Section~\ref{sec: discussion} discusses implications for industrial practice, and Section~\ref{sec: conclusion} concludes with reflections and directions for future work.

\begin{figure*}
    \centering
    \includegraphics[width=\linewidth]{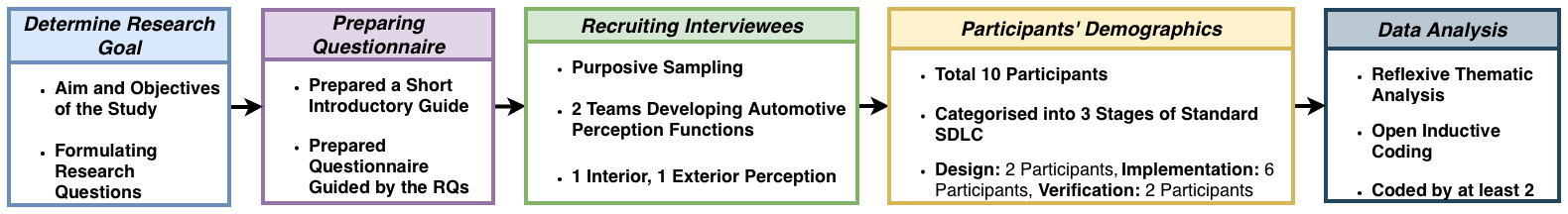}
    \caption{A graphical overview of the study.}
    \label{fig:study overview}
\end{figure*}

\section{Background and Related Work}
\label{sec: background}
Work on data leakage spans multiple research communities: reproducibility studies in ML, computer-vision methods for detecting near-duplicates and temporal overlap, and software-engineering research on how data-centered practices integrate into engineering lifecycles. Taken together, literature from these fields explain both the technical ways leakage shows up in perception systems and the operational reasons it persists in industry.

Reproducibility and benchmark audits make clear that leakage is a recurring, practical problem. Empirical studies show that small, hidden correlations between training and test sets can dramatically inflate reported performance; when testbeds are modified or when stricter data split policies are applied, performance drops reveal that models often exploit dataset idiosyncrasies rather than capture robust characteristics \cite{recht2019imagenet, kapoor2023leakage}. These findings motivate treating data leakage as an engineering hazard that may undermine the validity of offline evaluation and misleads downstream verification activities.

Computer vision research offers concrete detection approaches for image datasets. Techniques fall into a few pragmatic families: perceptual hashing for quick duplicate checks; embedding-driven retrieval using deep visual features for more tolerant similarity detection; and hybrid pipelines that combine local region matching with global descriptors to handle viewpoint and occlusion differences \cite{thyagharajan2021review, oquab2023dinov2, douze2025faiss}. Embedding-based pipelines, in particular, have become practical with self-supervised representations and scalable nearest-neighbour indices. They work well when metadata is noisy but require computational resources and careful thresholding to avoid false positives \cite{oquab2023dinov2, douze2025faiss, song2024train}. In automotive perception, these methods are important, because datasets commonly contain repeated routes, recurring environmental contexts, and sensor configurations that produce near-overlapping frames across captures.

Parallel to algorithmic tools, researchers in the field of software engineering also examined how engineering processes, tool chains, and organizational practices influence data quality. Sculley et al. introduced the notion of hidden technical debt to show that data dependencies, feature coupling, and under-documented preprocessing steps become operational liabilities as systems evolve \cite{sculley2015hidden}. Subsequent empirical case studies and practitioner-oriented reports emphasize similar points: dataset and model versioning, experiment tracking, and dataset lineage are unevenly adopted; teams often treat data as a transient artifact rather than a first-class engineered component \cite{amershi2019software, breck2017ml, serban2024software}.

Governance proposals and documentation artifacts aim to make data usage and evaluation transparent. “Datasheets for datasets” and “Model Cards” are recommended practices that record collection methods, intended uses, and known limitations; they make it easier for teams to reason about evaluation scope and potential data leakage \cite{gebru2021datasheets, mitchell2019model}. In principle, these artifacts combined with MLOps primitives (dataset versioning, sealed evaluation sets, reproducible preprocessing pipelines, etc.) provide a robust defense: they enable traceability and reproducible re-runs that can detect whether a reported result depends on contaminated data. In practice, however, adoption varies: some teams have solid experiment logging yet weak dataset provenance, while others control datasets but lack having immutable evaluation sets. That mismatch is a recurring theme in the literature \cite{amershi2019software, serban2024software}.

The automotive domain adds further constraints and considerations. Safety standards and verification regimes such as ISO26262 \cite{ISO26262} and safety of the intended functionality (SOTIF) shape how teams think about faults and hazards, but these standards were not designed for data-driven artifacts and offer limited prescriptive guidance on preventing dataset contamination \cite{borg2020safely}. Developers therefore operate at the intersection of two practices: rigorous, standards-driven verification and validation (V\&V) on one hand, and the exploratory data engineering realities of ML on the other. Several applied studies in automotive perception have proposed domain-aware mitigation such as route-aware splits, scenario mining, spatio-temporal separation, and metadata-based approaches to reduce the chance of overlap across splits \cite{babu2025d, heyn2023automotive}. These domain-specific techniques are effective, but they depend on high-quality metadata and require organizational commitment to run costly checks at scale.

Finally, despite this growing body of technical and process guidance, the literature lacks a focused, role-sensitive empirical account of how leakage is perceived and handled inside industrial teams across various roles. Prior studies provide scattered evidence that different roles approach data quality and validation concerns from distinct perspectives: ML engineers often emphasize algorithmic fixes and embedding-based checks \cite{kapoor2023leakage, song2024train}, verification engineers focus on reproducible evaluation and immutable testbeds \cite{borg2020safely, amershi2019software}, while system designers link data assumptions to ODD definitions and scenario coverage within safety-oriented development frameworks \cite{hoss2022review, de2024coverage}.

\section{Methodology}
\label{sec: methodology}
We have conducted an exploratory case study to explore the understanding and practices concerning data leakage in an automotive OEM. To achieve this, we performed semi-structured interviews with the ten participants from different development roles to elicit their own definitions, experiences, and mitigation strategies of data leakage and related risks in the development workflow. This approach prioritizes knowledge depth, contextual understanding, and interpretive validity over statistical generalization \cite{runeson2009guidelines, braun2006using, tong2007consolidated}. A detailed overview of the study is illustrated in Figure \ref{fig:study overview}.

\subsection{Case Study Context}
The company where this study was conducted develops a broad range of both interior and exterior perception functions as part of its automotive software systems. Exterior perception functions include object detection and lane recognition, while interior perception covers driver monitoring, occupant detection, and other in-cabin sensing applications.

In this context, the case refers to a set of perception-function development projects within an automotive OEM, representing typical industrial machine-learning pipelines that operate under established quality and reliability constraints. The case was selected because it provides a realistic instance of how ML-based perception is integrated into conventional automotive development governed by standards such as ISO 26262. The organization’s mature software development processes, combined with active ML deployment make it a representative example of current industrial practice. 
The study involved two perception teams within this organizational setting; details of participant roles and recruitment are provided in the next subsection.

\subsection{Data Collection}
We conducted interviews with ten practitioners from two development teams responsible for ML-enabled perception functions: one focused on interior perception and the other on exterior perception-related functions. To facilitate meaningful role-based comparisons, we classified participants into three categories:

\begin{itemize}
    \item \textbf{Design Phase (2 participants):} Engineers responsible for system designing and software  architecture and interfaces, ensuring design consistency and alignment with functional requirements. 
    \item \textbf{Implementation Phase (6 participants):} ML engineers, data scientists, and data engineer/architect involved in the preparation of the data sets, the development of models, and the relevant software components.
    \item \textbf{Verification/Testing Phase (2 participants):} Verification engineers responsible for system validation, testing pipelines, and quality assurance.
\end{itemize}

Participants were recruited via purposive sampling to ensure coverage of all relevant roles \cite{campbell2020purposive}. All the interviews were conducted through Microsoft Teams meetings. Table~\ref{tab:demographics} summarizes the demographics of the participants.

\begin{table}
    \centering
    \footnotesize
    \caption{Participant demographics (N=10) with their role in the development process and the number of years of experience within the current roles.}
    \label{tab:demographics}
    \begin{tabular}{c c c c}
        \hline
        \textbf{ID} & \textbf{Role} & \textbf{SDLC Category} & \textbf{Experience (Yrs.)} \\
        \hline
        P1 & System Design Engineer & Design & 1--1.5 \\
        P2 & System Design Engineer & Design & 1  \\
        P3 & Data Architect & Implementation & 2  \\
        P4 & Data Scientist & Implementation & 10+  \\
        P5 & ML Engineer & Implementation & 5 \\
        P6 & ML Engineer & Implementation & 4  \\
        P7 & ML Engineer & Implementation & 3  \\
        P8 & ML Test Engineer & Implementation & 4  \\
        P9 & Verification Engineer & Verification & 3  \\
        P10 & Verification Engineer & Verification & 3  \\
        \hline
    \end{tabular}
\end{table}

\begin{table*}
    \centering
    \caption{The list of interview questions mapped into the corresponding research questions.}
    \label{tab:questionnaire}
    \begin{tabular}{c p{13cm} c}
        \hline
        \textbf{Question No.} & \textbf{Questions} & \textbf{Related RQs} \\
        \hline
        Q1 & How would you define data leakage generally? How in the context of image data? & RQ1 \\
        Q2 & Do you think data leakage is a common problem in ML projects? Why or why not? & RQ1 \\
        Q3 & Do you consider data leakage when planning training and validation? & RQ1 \\
        Q4 & Have you come across any notable examples of data leakage (from research papers, case studies, or real-world projects)? & RQ1 \\
        Q5 & Have you ever encountered data leakage in your own work? If so, can you describe the situation? & RQ2 \\
        Q6 & What was the impact of the data leakage (e.g., overestimated performance, poor generalization)? & RQ2 \\
        Q7 & How was it discovered, and what steps were taken to resolve it? & RQ2 \\
        Q8 & What techniques or best practices do you use to detect data leakage in your ML pipelines and which phases of the development? & RQ3 \\
        Q9 & Do you use any tools or frameworks to check for data leakage? What strategy do you follow? & RQ3 \\
        Q10 & At what stages of the ML lifecycle do you think data leakage is most likely to occur? (e.g., data collection, preprocessing, model training, evaluation) & RQ3 \\
        Q11 &  What action do you usually take and/or prefer to take in the case of data leakage detection? & RQ3 \\
        Q12 & How do you prevent data leakage during feature engineering and data splitting? & RQ3 \\
        Q13 & What measures do you think teams should take to prevent data leakage? & RQ4 \\
        Q14 & How do you educate team members about the risks of data leakage? & RQ4 \\
        Q15 & If you could implement one industry-wide best practice to prevent data leakage, what would it be? & RQ4 \\
        \hline
    \end{tabular}
\end{table*}

The interview was designed in a semi-structured format, and the questionnaire consists of a combination of both open and closed questions. Table \ref{tab:questionnaire} presents the list of all the questions, and mapping of them to the corresponding RQs. Each interview was between 30–60 minutes long. The questions were grouped based on the research questions:

\begin{enumerate}
    \item \textbf{Definition and Knowledge:} Participants’ understanding of ``data leakage'' and its relevance to their role.  
    \item \textbf{Experience and Examples:} Specific examples or observations of data leakage, including perceived impact.  
    \item \textbf{Detection and Prevention:} Methods, tools, or processes used to detect, prevent, or mitigate leakage.
    \item \textbf{Recommendation Guidelines:} Guidelines and best practices recommended by the practitioners.
\end{enumerate}

The semi-structured format allowed follow-up questions on domain-specific examples and probing differences across SDLC categories. All interviews were conducted in accordance with the ethical principles of the ACM and IEEE codes of research conduct. Participation was entirely voluntary, and each participant provided informed consent before the interview. To ensure confidentiality, no personal or company-identifiable information was recorded or reported, and all transcripts were anonymized before analysis.

\subsection{Data Analysis}
We analyzed the interview data using reflexive thematic analysis following Braun and Clarke’s established guidelines \cite{braun2006using}. The process was iterative and interpretive. First, two of the authors familiarized themselves with the transcripts by reading them multiple times, taking notes, and writing short analytic notes about observations. Next, they performed open coding inductively, capturing meaningful excerpts and assigning initial labels using a shared coding sheet. These codes were then grouped and refined through axial coding to identify higher-level categories that corresponded to key stages of the SDLC, such as design, data preparation, modeling, integration, and verification, as well as to role-specific patterns across participants. Finally, theme labels were consolidated, relationships among them were examined, and visual maps were produced to clarify how each theme related to the research questions. The first two authors coded the transcripts sitting together over seven coding sessions and reconciled differences through discussion to strengthen interpretive consistency. We intentionally avoided automated topic modeling, since our goal was to preserve contextual richness and capture the nuances of practitioners’ reasoning rather than to perform statistical clustering.

We followed practical saturation guidelines: after approximately eight interviews, few new themes emerged, consistent with prior research indicating thematic saturation often arises within 6–12 interviews in homogeneous organizational contexts \cite{guest2006many, hennink2017code}. Thus, the sample provides sufficient depth for analytic generalization while acknowledging limits on statistical representativeness.

To ensure rigor, our study addressed the following:

\begin{itemize}
    \item \textbf{Triangulation:} Inclusion of design, implementation, and verification roles to capture cross-role perspectives on the topic.  
    \item \textbf{Researcher Triangulation:} Multiple coders with iterative consensus and synchronized discussions.  
    \item \textbf{Reflexivity:} Analysts maintained notes on preconceptions and domain positions.  
\end{itemize}

\section{Findings}
\label{sec: findings}
\begin{figure*}
    \centering
    \includegraphics[width=0.9\linewidth]{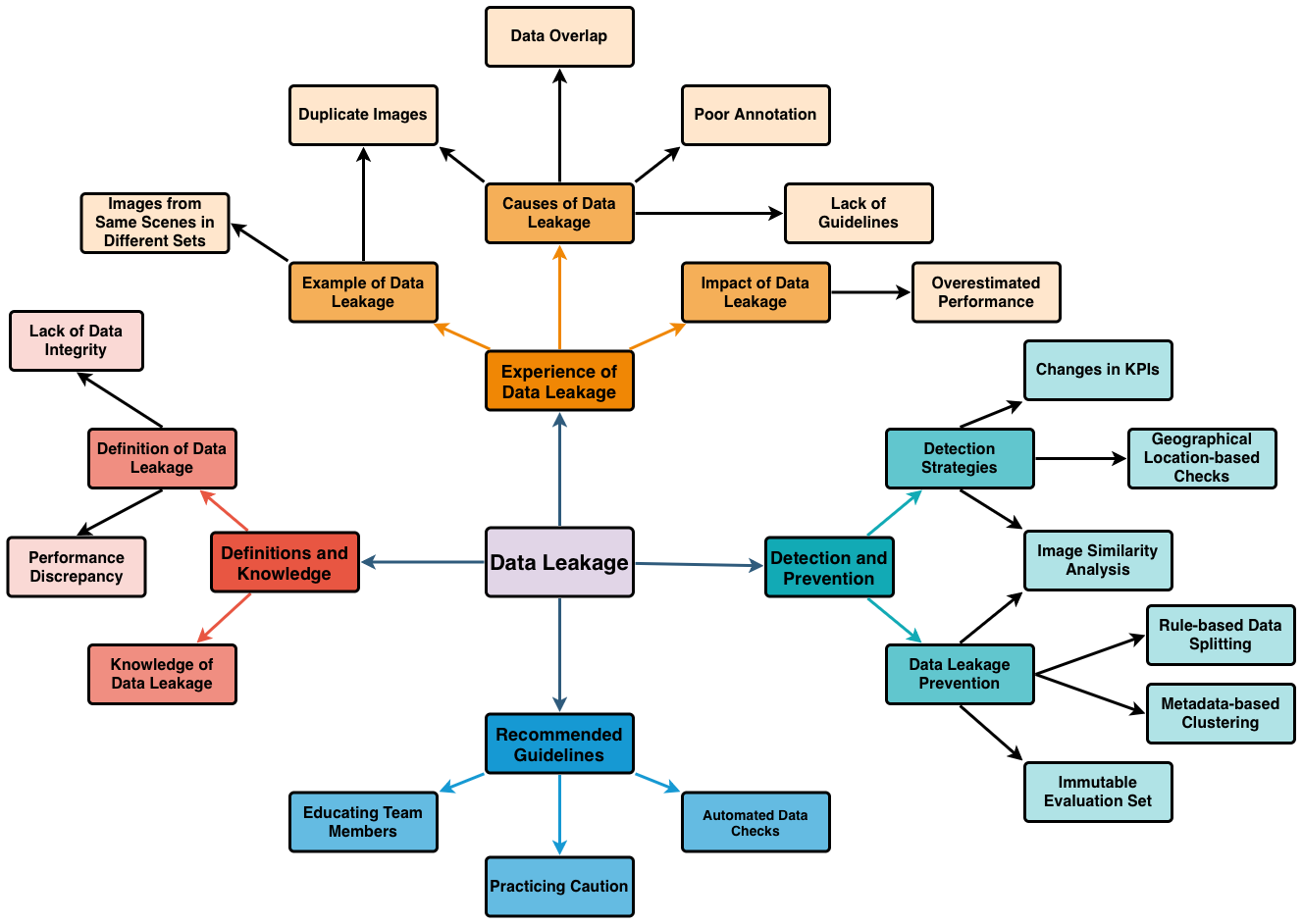}
    \caption{A mindmap of the main themes and sub-themes produced from the participants' responses and also motivated by the RQs. The leaf nodes contain the most frequent codes mentioned by the practitioners. Some codes were repeatedly mentioned while answering questions related to different themes, such as \textit{Duplicate Images} and \textit{Image Similarity Analysis}. \textit{Duplicate Images} were used to give examples, so it also came out as one of the mentioned causes of data leakage for obvious reasons. Practitioners also repeatedly mentioned \textit{Image Similarity Analysis}, which can be used to detect data leakage as well as to prevent it.}
    \label{fig:data leakage mindmap}
\end{figure*}

This section reports the empirical findings organized by the four research questions. We report the main themes that answer each RQ. The themes are shown in the mindmap presented in Figure \ref{fig:data leakage mindmap}. Where participant language is used, we prefer composite paraphrases to single-speaker quotes so as to preserve anonymity and to respect agreed reporting constraints.

\subsection{Definitions and Knowledge — RQ1}
Three concise observations summarize our answers to RQ1.

\paragraph{\textbf{Data Leakage Definition}}
During the interview, participants were asked to define data leakage according to their own understanding and using their own words. Regardless of the role in the development process, their way of defining data leakage was very similar. One participant (P2) from the design group defined data leakage as having some sort of invalidated data. Participants in the validation group also referred to data leakage, having the same data in both the train and validation datasets. 

However, the more technical roles from the implementation category used a few technical terms to give more broader definition. Participant P4 mentioned that even having very similar data in the evaluation can also be called data leakage. 
\begin{participantquote}
    Using data for evaluation that is either the same data which was used during training, or is very similar to the training data. --- \textbf{P4}
\end{participantquote}
P3 used different terminology, like data integrity, to give the definition of data leakage.
\begin{participantquote}
    Data leakage occurs when a cluster of data with an intended purpose cannot guarantee the integrity of that cluster. The three main intents that we handle here are training, evaluation, and test, and you cannot guarantee integrity when you have similar subjects present in two clusters. --- \textbf{P3}
\end{participantquote}
In essence, the responses can be synthesized to define data leakage as a phenomenon with having either same data in training and evaluation sets or having very similar data shared in both the sets. The participants were subsequently asked to define data leakage in the image data context, to see whether they tend to give the same definition. 8 out of the 10 participants defined it the same way as they defined data leakage in general. Two participants changed the wording of their definition a little. P5 added that having similar images in training and evaluation sets can also be called data leakage.
\begin{participantquote}
    When we have the same or similar images in the training and evaluation set, such as images from the same view, or same weather, or the same time. --- \textbf{P5}
\end{participantquote}
The other answer, by P3, was indirect, stating that the definition would be determined depending on the context and the data subjects. Only this definition mentions the performance discrepancy on the evaluation and test sets.
\begin{participantquote}
    Depends on the subjects and context, which can be determined by the discrepancy of KPIs on evaluation and test sets. --- \textbf{P3}
\end{participantquote}

\paragraph{\textbf{Knowledge across roles}}  
Knowledge and experience of data leakage are clustered around job responsibilities. ML-facing roles (ML engineers, data scientists, data architect) tended to describe leakage through technical mechanisms, representing their knowledge and experience depth of the subject. Design and Verification/Testing roles framed their level of knowledge and concerns about data leakage in terms of test scope, scenario representativeness, and verification artifacts. The participants were also asked whether they considered the risk of data leakage during their work or not. All job roles in the implementation phase replied affirmatively.
\begin{participantquote}
    Every single colleague we have working on machine learning is painfully aware of this. --- \textbf{P3}
\end{participantquote}
Despite participants from the design and verification roles being aware of data leakage, one system design engineer acknowledged that they did not consider data leakage. This is due to the fact that knowledge of data leakage is not required to perform their tasks, which is designing the system architecture, since data leakage is a problem that can arise during implementation only.
\subsection{Experience of Data Leakage - RQ2}
Participants were asked if they could remember any cases of data leakage during their regular work in the corresponding roles. 

\paragraph{\textbf{Example of data leakage experienced by practitioners}}
The design group and Verification group have mentioned that they have never come across any specific example of data leakage, as they have never encountered data leakage in their day-to-day work. However, all the participants in the \textit{Implementation} group have encountered data leakage either during their personal/academic projects development or during the development work in the industry. Participants who encountered this issue in the industry also mentioned that the issue was diagnosed in the feasibility or pilot run phase, and much prior to the actual product/project deployment.
\begin{participantquote}
    There was no severe consequence because eventually we noticed the problem before it continued developing, we were very far from production, and the leakage issue was stopped there. --- \textbf{P7}
\end{participantquote}

\paragraph{\textbf{Causes of data leakage}}
The participants who experienced data leakage were also asked to explain the situation and the possible reasons behind the cause of leakage. The majority of them acknowledged that the issue was caused by or during data splitting.
\begin{participantquote}
    When splitting the data, I was putting together all images collected by different vehicles and doing a random split, 70\% for the training and 30\% for validation. --- \textbf{P7}
\end{participantquote}
Other participants (\textbf{P4}, \textbf{P5}, \textbf{P6}, \textbf{P8}) also mentioned similar reasons behind the occurrence of data leakage. \textbf{P9} mentioned that the cause of data leakage was constantly re-splitting the data because, in that case, no valid dataset for testing is kept. Another participant interestingly mentioned that they intentionally made a leakage due to a lack of data in a pilot study, so that model was never released. \textbf{P3} mentioned noticing the presence/occurrence of data leakage in many research papers and projects done by other people, just because of the lack of proper knowledge about data leakage. Two participants explicitly mentioned at this stage that having similar images from the same scene across training and evaluation sets actually caused data leakage.
\begin{participantquote}
    We found that there has been driving data from the same location but different directions, which are semantically the same, but not in image space. --- \textbf{P5}
\end{participantquote}

\paragraph{\textbf{Impact of data leakage}}
The participants were also asked about the impact of data leakage perceived by them. Many of them agreed, saying the performance metric would be biased with overestimated values. 
\begin{participantquote}
    It was an unexpectedly high performance for scenarios where we don't have much data for those kinds, and we expected it to be kind of poor. --- \textbf{P6}
\end{participantquote}
To answer what steps were taken to resolve the data leakage issue, participants emphasized carefully splitting the data. They mentioned the importance of properly separating data from the same sequence and not using them for both training and validation.

\subsection{Detection and Prevention Practices - RQ3}
Practitioners described a pragmatic mix of technical checks and process controls to detect and/or prevent data leakage. 

\paragraph{\textbf{Data Leakage Detection}}
The participants' answers can be divided mainly into two groups. The first group, which includes an ML engineer from the implementation stage and a verification engineer from the verification stage, agrees on noting the model performance as an indicator of data leakage. When the model performance indicator looks unreasonably high (i.e., too good to be true), it might be a sign of data leakage.
\begin{participantquote}
    When it's (model performance) too good to be true, it's usually something you should doubt and recheck --- \textbf{P5}
\end{participantquote}
The larger group of participants' responses mentioned image similarity analysis and geographical location-based checks (e.g., GPS heatmap, metadata-based clustering) to detect the presence of data leakage. Since the participants work on perception function development for both interior and exterior perception, they work mainly with image data, associated with other sensor data from LiDAR and Radar.
\begin{participantquote}
    We also use image similarity checks to see if we have almost identical frames. --- \textbf{P6}
\end{participantquote}
One participant answered differently, instead of mentioning a straightforward way of detection. The participant thinks that one has to have proper knowledge about the function they are developing and its operational design domain (ODD). Since the definition varies based on the scope of the function or task, the way of data leakage detection will also vary as per that participant.

\paragraph{\textbf{Strategy/method for leakage detection}}
We also asked participants about strategies they follow or would like to suggest for data leakage detection. Participants mentioned using pre-trained models like CLIP \cite{hafner2021clip} for feature extraction from the images. The extracted features are then used to compute similarity by using simply Euclidean distance or Cosine similarity to group the similar images into the same cluster (train or evaluation). Another method they brought into the discussion is metadata-based clustering. Grouping images collected from the same geographical area, under the same lighting and/or weather conditions, etc., into the same cluster.
\begin{participantquote}
    (...) using GPS data if available to cluster data collected from the same area (...) --- \textbf{P6}
\end{participantquote}

\paragraph{\textbf{Steps when data leakage could occur}}
Most of the participants think that data leakage can occur during data splitting (i.e., when they separate the training and evaluation data). One of the verification engineers, similar to an ML engineer, said that it can happen anywhere between data collection and model training.
\begin{participantquote}
    Of course, splitting. This is the place where you separate the data, right? So, if you did it wrong, then it (data leakage) actually occurs. --- \textbf{P4}
\end{participantquote}
One ML engineer pointed to the initial steps when data is being collected and annotated as the step when data leakage can happen. The participant thinks that similar data should not be added to the dataset simply because a representation of the data point is already present in the existing dataset.
\begin{participantquote}
    I think it's in step zero when we choose what to collect and annotate (...) when we do data preparation, data creation, we don't even want to add the same data because we already have it. --- \textbf{P5}
\end{participantquote}

\paragraph{\textbf{Data leakage prevention}}
Practitioners across roles were also able to suggest what preventive measures can be taken against data leakage. System designers think that there should be some additional data requirements to ensure data integrity across the development stages. The requirements must include data representativeness and coverage of all possible scenarios during data collection. Practitioners in the implementation stage have overlapping thoughts on the prevention measures. ML engineers repeatedly mentioned the importance of applying rule-based data splitting and applying one additional step for similarity checks.
\begin{participantquote}
    The simplest approach is to have simple logic in the splitting, like time and location (...) --- \textbf{P5}
\end{participantquote}
One practitioner specifically mentioned the importance of keeping the immutable evaluation dataset from completely different locations than the training data. The data should follow a uniform distribution in terms of metadata while being isolated and immutable, the participant added. The data architect, however, still places emphasis on the importance of being cautious and having task-specific knowledge in addition to applying rule-based data splitting. The practitioners from the verification group tend to focus only on the data representativeness and good coverage of every possible scenario in order to shield the model from being under-performing or over-performing on a specific task.

\subsection{Recommended Guidelines from Practitioners - RQ4}
When it comes to educating team members about the risks that might be posed by data leakage and also to recommend industry-wide guidelines, the practitioners responded differently.

\paragraph{\textbf{Educating team members of data leakage risks}}  
Practitioners, particularly from the implementation group, emphasized more on explaining the topic hands-on; for example, by showing and explaining red flags and real example cases. One participant explicitly mentioned being cautious in each step and maintaining reports or version control of datasets while new data is added.
\begin{participantquote}
    I would like to raise awareness so that people know the things that are not good and they deserve to look into (...) There should be a report available stating the current red flags (...) --- \textbf{P8}
\end{participantquote}
A few participants refused to specify any particular guideline because they believe data leakage situations can be different for different tasks and the education has to be task-specific, otherwise incorrect theories might be applied to avoid data leakage at the end by new team members. A verification engineer has a different thought and wants to educate colleagues about data leakage by teaching them to notice if there is any performance discrepancy of the model in real-world scenarios.

\paragraph{\textbf{Practitioners' recommendations and guidelines}}  
The recommendations made by the practitioners reflect their own way of preventing data leakage in most cases. A frequent suggestion was to practice caution and be well aware from the data collection stage, follow standard ways of data splitting. The importance of knowing the specific task well to be able to prevent data leakage is repeated again as a recommended guideline. Most of the implementation-related roles mentioned the necessity of having a preserved evaluation set along with the placement of automated similarity checking steps before adding new data to the existing datasets.
\begin{participantquote}
    (...) When we add new data, we have to have a way to check (for data leakage) (...) --- \textbf{P5}
\end{participantquote}
Another participant from the same (Implementation) group also mentioned the necessity of having image embedding-based similarity checks. One of the verification engineers went more specific and recommended being cautious and following standard data splitting ways that serve the specific task better, and with the possible inclusion of appropriate metadata.
\begin{participantquote}
    My suggestion is, we should have a standard to follow when we split the data for training and validation. --- \textbf{P10}
\end{participantquote}
Since the design-specific roles are not responsible for implementation, they did not have any specific recommendations for the implementation-specific group of practitioners.

\section{Discussion}
\label{sec: discussion}
Our findings reveal that practitioners across different roles within the automotive function development lifecycle interpret and manage data leakage through diverse but complementary lenses. This section discusses what these findings imply for both research and industry practice, reflecting on awareness, experience, detection, and prevention of data leakage risks. We connect the role-based perceptions observed in this study to broader discussions on reliable AI engineering.

Overall, the interviews show that the practitioners have knowledge of \emph{data leakage}, but its interpretation is shaped by role boundaries. ML engineers and data scientists frame leakage as a statistical or procedural flaw in dataset preparation and model validation. In contrast, system designers and verification engineers treat it as an issue of representativeness or scenario completeness rather than data overlap. This asymmetry suggests that while knowledge exists, a shared operational definition is lacking, making communication across SDLC stages inconsistent. 
In industrial settings, such gaps can delay issue detection and lead to quality drift, where a leakage problem introduced early may only become apparent during later validation stages.

% \begin{tcolorbox}[colback=blue!5!white,colframe=blue!30!white,title=\textbf{Answer to RQ1:}]
%     \emph{Data leakage is commonly understood across roles as overlap or contamination between training and evaluation datasets}, but the degree of technical depth varies by role. Implementation roles describe leakage in more technical and operational terms, while design and verification roles approach it through representativeness and test coverage.
% \end{tcolorbox}

\begin{quote}
\noindent\textbf{Answer to RQ1:}

\emph{Data leakage is commonly understood across roles as overlap or contamination between training and evaluation datasets, but the degree of technical depth varies by role. Implementation roles describe leakage in more technical and operational terms, while design and verification roles approach it through representativeness and test coverage.}
\end{quote}

From a research perspective, our findings point to a gap in how ML reliability practices are typically conceptualized. Data leakage is often framed as a technical issue to be addressed through tools, validation strategies, or dataset management rules. However, the interviews indicate that practitioners’ understanding of leakage is shaped by their role, their responsibilities, and the information available at their stage of the development process. In this sense, managing leakage requires not only technical safeguards but also coordination, shared interpretations, and clarity about responsibilities across teams—elements that are inherently socio-technical. Strengthening cross-role communication, establishing common definitions, and improving shared documentation could therefore support more consistent detection and prevention of leakage throughout the ML development lifecycle.

Experiences with leakage were limited to those directly handling data, typically within the implementation group. The problems surfaced mainly in either non-production contexts, academic projects, feasibility studies, or early pilot tests, suggesting that organizational containment mechanisms are effective once systems move toward integration. However, these cases also expose the importance of data management workflows. The use of similar or temporally adjacent images in both training and evaluation datasets is a recurring issue in perception systems, particularly when data originates from continuous data collection streams. Hence, practitioners learning from past experience should practice extra caution while developing models using such data streams.

% \begin{tcolorbox}[colback=blue!5!white,colframe=blue!30!white,title=\textbf{Answer to RQ2:}]
%     Practitioners encountered data leakage during experimental or pilot phases rather than deployment. The causes are typically tied to data splitting practices and insufficient awareness of sequence- or scene-level similarity.
% \end{tcolorbox} 

\begin{quote}
\noindent\textbf{Answer to RQ2:}

\emph{Practitioners encountered data leakage during experimental or pilot phases rather than deployment. The causes are typically tied to data splitting practices and insufficient awareness of sequence- or scene-level similarity.}
\end{quote}

This finding resonates with industry-wide concerns about the fragility of ML pipelines, where seemingly minor data-handling mistakes can produce significant metric inflation. Therefore, improved model traceability and dataset versioning, particularly at the image-sequence and metadata levels, emerge as essential technical debt items for future process maturity.

Practitioners approach leakage detection through a blend of generic considerations and lightweight tooling. Model performance anomalies, which are “too good to be true” results, serve as a common early warning sign. Some engineers incorporate similarity metrics or metadata clustering to detect near-duplicate samples, particularly for image data.

% \begin{tcolorbox}[colback=blue!5!white,colframe=blue!30!white,title=\textbf{Answer to RQ3:}]
%     Detection relies on a mix of performance monitoring and data similarity checks, while prevention focuses on structured data splitting and contextual domain knowledge. Current approaches are role-specific rather than broadly systematic.
% \end{tcolorbox}

\begin{quote}
\noindent\textbf{Answer to RQ3:}

\emph{Detection relies on a mix of performance monitoring and data similarity checks, while prevention focuses on structured data splitting and contextual domain knowledge. Current approaches are role-specific rather than broadly systematic.}
\end{quote}

Preventive measures, in turn, are guided by experience rather than formal policy. ML engineers emphasize rule-based splitting (e.g., temporal, geographical, or sequence separation), while design and verification roles prioritize representativeness of test data. The challenge lies in ensuring both statistical independence and functional coverage. For practitioners, embedding these principles into CI/CD pipelines, with automated data similarity checks, would improve process robustness without imposing excessive manual burden.

Participants unanimously acknowledged the need for greater and more structured knowledge sharing around data leakage. Their suggested guidelines emphasize experiential learning through real leakage examples and “red flag” indicators rather than abstract rules. Implementation-focused roles advocate for dataset reporting, version control, and immutable evaluation sets, whereas system design roles call for clear data requirements during early stages.

% \begin{tcolorbox}[colback=blue!5!white,colframe=blue!30!white,title=\textbf{Answer to RQ4:}]
%     Practitioners recommend knowledge sharing by demonstrating leakage examples, maintaining dataset traceability, and integrating rule-based similarity checks. They view leakage prevention as a cross-role responsibility requiring continuous education.
% \end{tcolorbox}

\begin{quote}
\noindent\textbf{Answer to RQ4:}
    
\emph{Practitioners recommend knowledge sharing by demonstrating leakage examples, maintaining dataset traceability, and integrating rule-based similarity checks. They view leakage prevention as a cross-role responsibility requiring continuous education.}
\end{quote}

The convergence on the importance of educating team members about data leakage underscores that leakage prevention cannot rely solely on technical tools; it requires continuous knowledge transfer to other team members. Organizations developing ML-based automotive functions could benefit from treating leakage awareness as part of their ML safety culture, similar to functional safety or cybersecurity training, embedding these concepts into design reviews, quality checklists, and onboarding processes.

Finally, reflecting across roles, our findings suggest that data leakage tends to emerge at points where responsibilities and assumptions intersect. Participants described different expectations about data provenance, dataset independence, and scenario similarity, and these expectations were not always aligned across teams. Such misalignments can make leakage more difficult to detect until later stages of development, even when each team individually follows established practices. Strengthening cross-role alignment—for example, through shared data documentation or clearer traceability of dataset transformations—may therefore help organizations identify potential leakage sources earlier in the development process.

\section{Threats to Validity}
\label{sec: threats to validity}
Following the guidance of Runeson and Höst \cite{runeson2009guidelines}, we discuss threats to validity under four categories: construct, internal, external, and reliability validity.

\subsection{Construct Validity}
Construct validity concerns whether the study accurately captures the concept it intends to investigate. In this case, how practitioners in automotive perception function development understand and manage data leakage. Because “data leakage” has no single operational definition across the ML lifecycle and the community, there is an inherent risk that interview questions or participants’ interpretations might vary by role or prior exposure. To mitigate this, we used open-ended prompts and allowed participants to express their own understanding before introducing any formal framing. The semi-structured format helped us balance structure with flexibility, allowing follow-ups to clarify role-specific meanings. Moreover, triangulating responses across three SDLC groups (design, implementation, verification) strengthened the construct representation by exposing overlapping and divergent interpretations.

\subsection{Internal Validity}
Internal validity relates to the credibility of the relationships inferred from the data. As in any interview-based study, the researcher's interpretation could bias coding or theme formulation. To address this, two analysts coded the transcripts through consecutive meetings and repeatedly reconciled differences through active discussion, reducing idiosyncratic judgments. Reflexive notes were maintained to track evolving assumptions and interpretive shifts. Another potential bias stems from role familiarity—participants may have emphasized practices viewed positively within their function or organization. To limit this effect, the interviewer explicitly clarified that the study sought reflective insights rather than evaluations of compliance or performance. Furthermore, inclusion of verification and system design participants, who interact with ML outputs but are not directly responsible for model training and development, offered contrasting perspectives that helped contextualize implementation-oriented claims.

\subsection{External Validity}
External validity concerns the generalizability or transferability of the findings. Our results are drawn from a single industrial context and a relatively small sample (ten participants), which naturally limits broad generalization. Nevertheless, the focus on role categories rather than organizational identity enhances analytical generalization: system designers, ML engineers, and verification specialists exist across most perception development pipelines. Thus, while specific tool references or workflows may vary, the observed cross-role gaps and differing conceptions of leakage are likely transferable to other ML projects in automotive or adjacent domains. To support reader assessment of transferability, we provide detailed participant demographics and clearly report the development stage each theme relates to.

\subsection{Reliability}
Reliability addresses the consistency and transparency of the research process. All interviews followed the same semi-structured guide, and interview questionnaire and notes, coding sheets, and theme definitions were versioned and archived. Two researchers performed coding and jointly validated the final themes. However, complete replication is limited by the interpretive nature of qualitative analysis and organizational confidentiality constraints, which restrict public sharing of transcripts. To enhance transparency, the analytic steps, data sources, and coding rationale are described in detail in Section~\ref{sec: methodology}.

\section{Conclusion and Future Work}
\label{sec: conclusion}
This study explored how practitioners across different software development lifecycle roles within the automotive perception domain interpret and address the problem of data leakage in machine learning workflows. Drawing on interviews with ten practitioners from \textit{design}, \textit{implementation}, and \textit{verification} roles, we found that data leakage is understood through distinct yet complementary lenses shaped by practitioners’ responsibilities and domain priorities. \textit{Implementation} roles tend to frame it as a technical artifact of data handling, while \textit{design} and \textit{verification} roles associate it with scenario coverage and representativeness. These differing perspectives highlight that leakage is a socio-technical coordination challenge distributed across the ML development pipeline.

Our findings contribute to a more nuanced understanding of how knowledge and responsibility for leakage are distributed in real-world ML engineering teams. They underscore the need to move beyond tool-centric solutions toward practices that promote shared understanding, cross-role communication, and traceable data management. Treating leakage control as a systemic property rather than an isolated data engineering task may help organizations achieve higher reliability in ML applications.

From a research standpoint, this work extends current discussions in software engineering for AI (SE4AI) by revealing how organizational boundaries shape the perception and management of data-related risks. While existing literature largely focuses on algorithmic or validation techniques, our results suggest that everyday engineering practices and communication routines play an equally decisive role in preventing data leakage from propagating through ML pipelines.

A central insight emerging from our study is that data leakage does not admit a single, universal definition. What constitutes leakage depends on the task, the model architecture, the data sources, and the overall application context. This contextual variability means that there is no straightforward, one-size-fits-all tool for detecting leakage. Instead, leakage should be viewed as a set of context-dependent failure modes that manifest differently across roles and stages of the ML pipeline, making systematic mitigation possible only when technical safeguards are complemented by a shared, cross-role conceptual understanding of what data leakage entails.

This study opens several directions for both research and industrial exploration. First, larger-scale, multi-organizational studies could help verify whether the observed role-based patterns generalize across domains beyond automotive perception. Second, longitudinal or ethnographic methods could capture how knowledge of data leakage evolves as teams adopt new tools and processes. Third, integrating empirical insights into model governance frameworks—linking dataset versioning, model lineage, and design rationale—could enable a more auditable and transparent ML development process. Finally, there is an opportunity to translate these findings into actionable engineering guidelines or lightweight checklists that complement existing safety and quality standards such as ISO~26262 and SOTIF, ensuring that data leakage prevention becomes a recognized component of AI reliability in practice.

\bibliographystyle{ACM-Reference-Format}
\bibliography{sample-base}
\end{document}